\documentclass{camera}
\usepackage{amsmath}
\usepackage{graphicx}  
\newcommand{\beq} {\begin{eqnarray}}
\newcommand{\eeq} {\end{eqnarray}}

\newcommand{\be}{\mbox{$^9$Be}}
\newcommand{\he}{\mbox{$^6$He}}

\newcommand{\aan}{\mbox{$\alpha+\alpha+n$}}
\newcommand{\bepb}{\mbox{$^9$Be+$^{208}$Pb}}
\newcommand{\ben}{\mbox{$^8{\rm Be}+n$}}
\newcommand{\ahe}{\mbox{$^5{\rm He}+\alpha$}}

\begin{document}
\title{A consistent four-body CDCC model of low-energy reactions: Application to $\bepb$}

\author{M. S. Hussein$^{1,2,3}$, P. Descouvemont$^{4}$ \and L. F. Canto$^{5,6}$}

\organization{$^1$ Departamento de F\'isica Matem\'atica, Instituto de F\'isica, Universidade de S\~ao Paulo, C.P. 66318, 05314-970, S\~ao Paulo, SP, Brazil\\
$^2$ Instituto de Estudos Avan\c{c}ados, Universidade de S\~ao Paulo, C.P. 72012, 05508-970, 
S\~ao Paulo, SP, Brazil\\
$^3$ Departamento de F\'{i}sica, Instituto Tecnol\'{o}gico de Aeron\'{a}utica, CTA, S\~{a}o Jos\'{e} dos Campos, S\~ao Paulo, SP, Brazil\\
$^4$ Physique Nucl\'eaire Th\'eorique et Physique Math\'ematique, C.P. 229,
Universit\'e Libre de Bruxelles (ULB), B 1050 Brussels, Belgium\\
$^5$ Instituto de F\'isica, Universidade Federal do Rio de Janeiro, C.P. 68528, 21941-972 Rio de Janeiro, RJ, Brazil\\
$^6$ Instituto de F\'isica, Universidade Federal Fluminense, Av. Gal. Milton Tavares de Souza s/n, Niter\'{o}i, RJ, Brazil} 

\maketitle

\begin{abstract}
We investigate the $\bepb$ elastic scattering, breakup and fusion 
at energies around the Coulomb barrier. The three processes are described simultaneously, with
identical conditions of calculations. 
The $^9$Be nucleus is defined in an $\aan$ three-body model, using the hyperspherical coordinate
method. We first analyze spectroscopic properties of $^9$Be, and show that the model provides
a fairly good description of the low-lying states. The scattering with $^{208}$Pb is then studied with the
Continuum Discretized Coupled Channel (CDCC) method, where the $\aan$ continuum is approximated by a
discrete number of pseudostates.  Optical potentials for the 
$\alpha+^{208}$Pb and $n+^{208}$Pb systems are taken from the literature. We present 
elastic-scattering and fusion cross sections at different energies. 
\end{abstract}

\section{Introduction}

Many experiments have been performed with the $\be$ nucleus, used as a 
target or as a projectile \cite{KAK09}.  Although $\be$ is stable, it presents a 
Borromean structure, as the well known halo nucleus $\he$.  None of the 
two-body subsystems $\alpha+n$ or $\alpha+\alpha$ is bound in $\be$,  
which has important consequences on the theoretical description of 
this nucleus.  Precise wave functions must include the three-body nature 
of $\be$.  The hyperspherical formalism \cite{ZDF93} is an ideal tool to 
describe three-body Borromean systems, as it does not assume a specific 
two-body structure, and considers the three particles $\alpha+\alpha+n$ on 
an equal footing.  

In the present work \cite{DDC15}, we aim at investigating $\be$ scattering and fusion on a heavy
target.  The reaction framework is the Continuum Discretized Coupled 
Channel (CDCC) method (see Ref.\ \cite{YMM12} for a recent review), which is well adapted 
to weakly bound projectiles since it allows to include breakup channels.  
Going from two-body projectiles (such as d=p+n or $^7$Li=$\alpha$+t) to three-body 
projectiles, however, strongly increases the complexity of the calculations, 
even if both options eventually end up with a standard 
coupled-channel system.  

Many data have been obtained for $\bepb$ elastic scattering 
\cite{WFC04,PJM11} and fusion 
\cite{DHB99,DGH04}.  These experimental data provide a good opportunity to test 
$\be$ wave functions. No assumption should be made about the cluster
structure, and $\alpha+^{208}$Pb as well as $n+^{208}$Pb optical 
potentials are available in the literature.

\section{Three-body model of $\be$}
\label{sec2}
The determination of the $\be$ wave functions is the first step for 
the $\bepb$ CDCC calculation.  
For a three-body system, the Hamiltonian is given by
\beq
H_0=\sum_{i=1}^3 \frac{\pmb{p}^2_i}{2m_i}+\sum_{i<j=1}^3 V_{ij}
(\pmb{r}_i-\pmb{r}_j),
\label{eq1}
\eeq
where $\pmb{r}_i$ and $\pmb{p}_i$ are the space and momentum coordinates 
of the three particles with masses $m_i$, and $V_{ij}$ a potential between nuclei $i$ and $j$.  
For the $\alpha+\alpha$ interaction, we use the deep potential of 
Buck {\sl et al.} \cite{BFW77}.  The $\alpha$+n interaction is taken from 
Kanada {\sl et al.} \cite{KKN79}.  Both (real) potentials accurately reproduce 
the $\alpha+\alpha$ and $\alpha$+nucleon phase shifts over a wide 
energy range.  

With these potentials, the $\be$ ground state is too bound 
($-3.12$ MeV, while the experimental value is $-1.57$ MeV with respect 
to the $\aan$ threshold).  We have therefore introduced a phenomenological three-body force which
reproduces the experimental 
binding energy of the $3/2^-$ ground state. 
The r.m.s. radius, the quadrupole moment, and the magnetic moment are $\sqrt{r^2}=2.36$ fm,
$Q(3/2^-)=4.96$ $e$.fm$^2$, and $\mu =-1.33\ \mu_N$, respectively. These values are in
fair agreement with experiment ($2.45\pm 0.01$ fm, $5.29\pm 0.04\ e$.fm$^2$ and $-1.18\ \mu_N$,
respectively) 

\section{The CDCC method}
\label{sec3}
\begin{figure}[h]
	\centerline{\includegraphics[width=.5\textwidth]{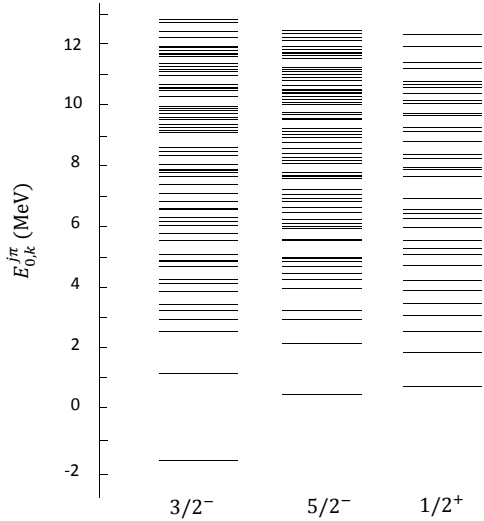}}
	\caption{Pseudostate energies (defined from the $\aan$ threshold) for the dominant
		partial waves $j=3/2^-, 5/2^-, 1/2^+$.}
	\label{fig1}
\end{figure}

We present here a brief outline of the CDCC method, and we refer to 
Ref.~\cite{RAG08} for specificities of three-body projectiles.  
The CDCC method is based on approximate solutions of the projectile Hamiltonian (\ref{eq1})
\beq
H_0 \Phi^{jm\pi}_k=E^{j\pi}_{0,k}\Phi^{jm\pi}_k,
\label{eq13}
\eeq
where $k$ are the excitation levels in partial wave $j\pi$.  
Solutions with $E^{j\pi}_{0,k}<0$ correspond to bound states of 
the projectile, whereas $E^{j\pi}_{0,k}>0$ correspond to narrow 
resonances or to approximations of the three-body continuum.  
These states cannot be associated with physical states, 
and are referred to as pseudostates. The $\be$ spectrum used
in the present model is shown in Fig. \ref{fig1}.

The Hamiltonian of the projectile + target system is written as
\beq
H(\pmb{R},\pmb{x},\pmb{y})=H_0(\pmb{x},\pmb{y})-\frac{\hbar^2}{2\mu_{PT}}
\Delta_R +V(\pmb{R},\pmb{x},\pmb{y})
\label{eq14}
\eeq
where $\mu_{PT}$ is the reduced mass of the system, and $\pmb{R}$ the relative coordinate. 
The Jacobi coordinates $\pmb{x}$ and $\pmb{y}$ are proportional to 
$\pmb{r}_{\rm Be-n}$ and $\pmb{r}_{\alpha-\alpha}$, 
respectively. The potential term reads
\beq
V(\pmb{R},\pmb{x},\pmb{y})=V_{1t}(\pmb{R},\pmb{y})
+V_{2t}(\pmb{R},\pmb{x},\pmb{y})+V_{3t}(\pmb{R},\pmb{x},\pmb{y})
\label{eq15}
\eeq
where the three components $V_{it}$ are optical potentials between 
fragment $i$ and the target.

In order to solve the Schr\"{o}dinger equation associated with 
(\ref{eq14}), the total wave function with angular momentum $J$ and parity $\Pi$ is expanded as
\beq
\Psi^{JM\Pi}_{(\omega)}(\pmb{R},\pmb{x},\pmb{y})=\sum_{j\pi kL}\,
\varphi^{JM\Pi}_{j\pi kL}(\Omega_R,\pmb{x},\pmb{y})
g^{J\Pi}_{jkL(\omega)}(R),
\label{eq16}
\eeq
where $\omega$ is the entrance channel, and where the channel wave functions are given by
\beq
\varphi^{JM\Pi}_{j\pi kL}(\Omega_R,\pmb{x},\pmb{y})=i^L
\biggl[\Phi^{j\pi}_k(\pmb{x},\pmb{y})\otimes
Y_L(\Omega_R)
\biggr]^{JM}.
\label{eq17}
\eeq
The radial functions $g^{J\Pi}_{c}(R)$ (we use the notation $c=(j,\pi,k,L)$) 
are obtained from the system
\beq
\biggl[ -\frac{\hbar^2}{2\mu_{PT}} \frac{d^2}{dR^2} +E_{c}-E \biggr] 
g^{J\Pi}_{c(\omega)}(R) 
+\sum_{c'}V^{J\Pi}_{c,c'}(R) g^{J\Pi}_{c'(\omega)}(R)=0,
\label{eq18}
\eeq
where $E$ is the c.m. energy, and where the coupling potentials are given by matrix elements
\beq
V^{J\Pi}_{c,c'}(R)=\langle \varphi^{JM\pi}_{c} \vert V \vert 
\varphi^{JM\pi}_{c'} \rangle
+\frac{\hbar^2}{2\mu_{PT}} \frac{L(L+1)}{R^2} \delta_{c c'} .
\label{eq19}
\eeq
As the fragment-target potentials (\ref{eq15}) are optical potentials, matrix elements (\ref{eq19}) contain
a real and an imaginary parts (see Ref. \cite{DDC15} for detail).
We solve Eq.\ (\ref{eq18}) with the $R$-matrix theory \cite{DB10}.

\section{Application to the $\bepb$ system}
\label{sec4}
\subsection{Elastic scattering}

\begin{figure}[h]
	\centerline{\includegraphics[width=.6\textwidth]{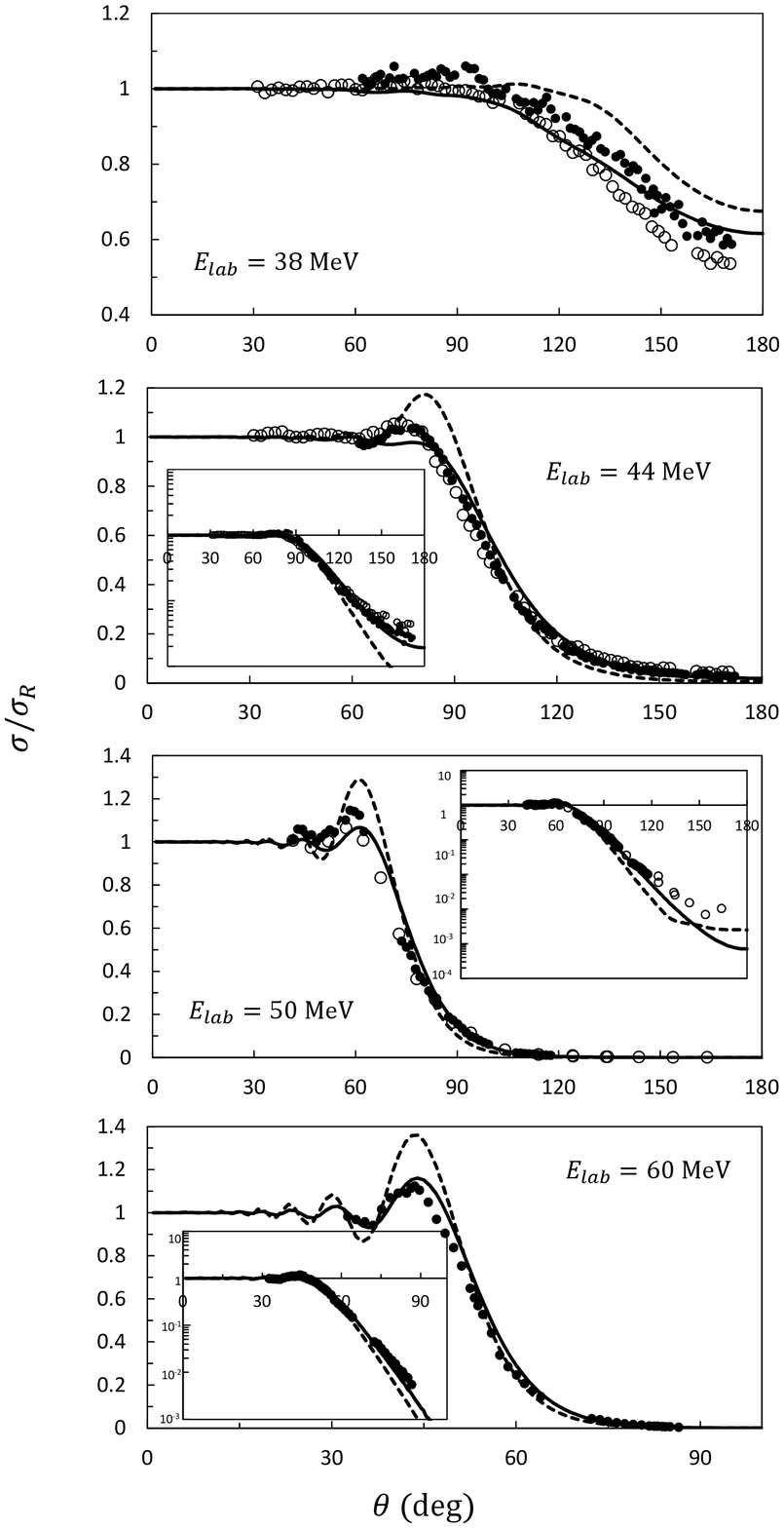}}
	\caption{$\bepb$ elastic cross sections (divided by the Rutherford cross section)
		at two $\be$ laboratory energies, and for different sets of $\be$ partial waves. 
		The experimental data are taken from
		Ref.~\cite{WFC04} (filled circles) and Ref.~\cite{YZJ10} (open circles).}
	\label{fig2}
\end{figure}

The calculations are performed with the $j^{\pi}=3/2^{\pm},5/2^{\pm},1/2^{\pm}$ partial waves on $\be$, and
the cutoff energy is 12.5 MeV.
Figure \ref{fig2} shows $\bepb$ elastic cross sections at two typical energies 
(the data are taken from Refs.\ \cite{WFC04,YZJ10}).  In contrast with 
optical-model calculations \cite{WFC04}, no renormalization of the potential 
is needed.  The absorption is simulated by the imaginary parts of the 
$\alpha-^{208}$Pb and n$-^{208}$Pb interactions, and by the $\alpha+\alpha+n$ 
discretized continuum.  
Calculations involving only the $3/2^-$ ground state are shown as dashed lines, 
and the full calculations as solid lines.  
From Fig. \ref{fig2}, it appears that continuum couplings 
are more important at low energies.  At $E=38$ MeV, the single-channel 
calculation, limited to the $\be$ ground state  is significantly 
different from the data.  This confirms the conclusion of a previous work \cite{DD12b} 
which suggests that continuum couplings are more important at low energies.

\subsection{Fusion}
The $\bepb$ fusion reaction has been studied theoretically by several groups
(see Ref.~\cite{JPK14} and references therein). Data about the various mechanisms (total TF,
complete CF and incomplete ICF fusion) are also available in the literature 
\cite{DHB99,DGH04,DHS10}.

\begin{figure}[h]
	\centerline{\includegraphics[width=.6\textwidth]{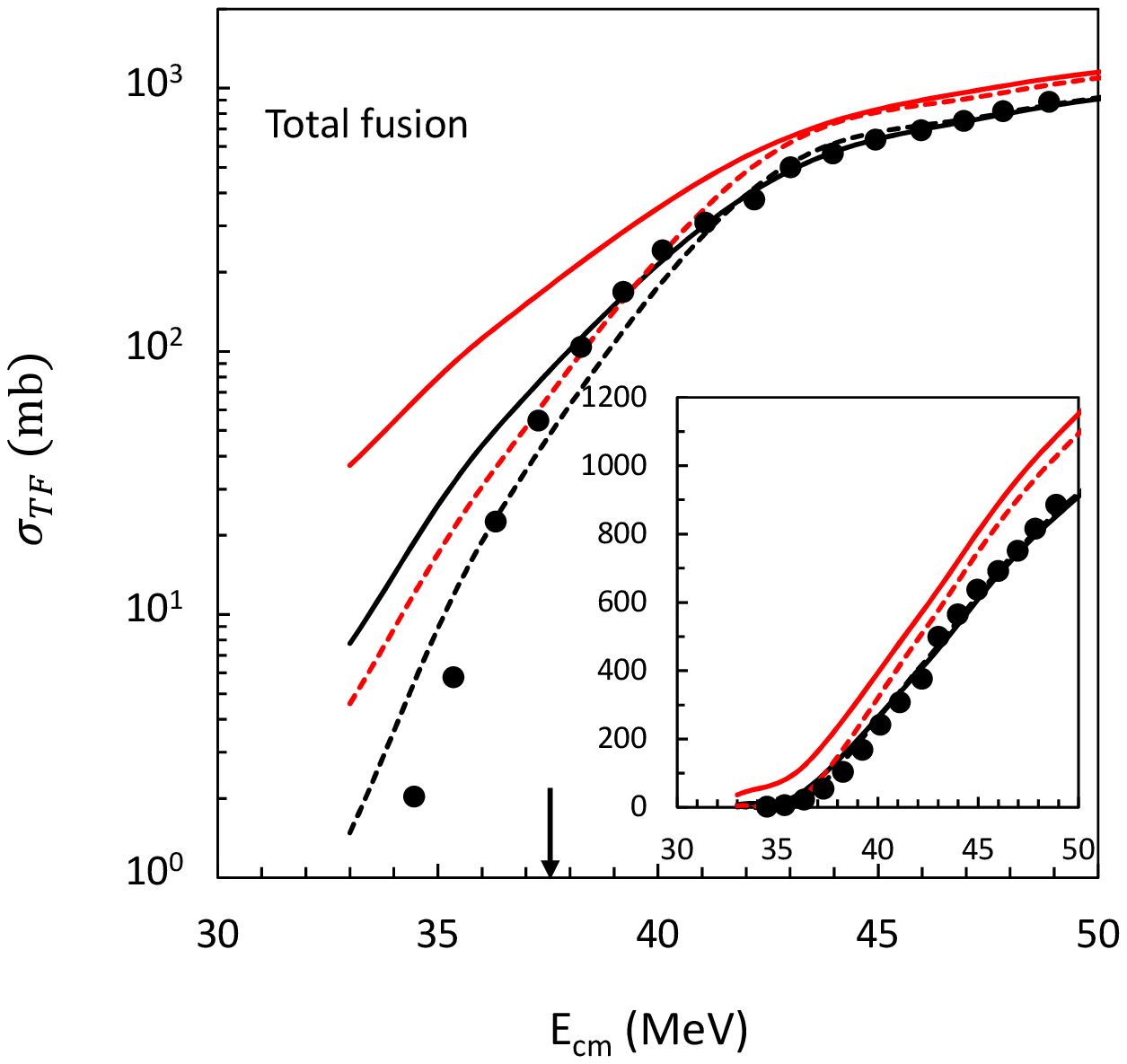}}
	\caption{Total fusion cross sections with (black lines) and
		without (red lines) removal of the neutron
		capture. The dashed lines represent the single-channel calculations, and the solid lines represent 
		the full calculations. 
		The Coulomb barrier energy $V_B$ is indicated by a vertical arrow.}
	\label{fig3}
\end{figure}

The total fusion cross section essentially involves two contributions: the complete fusion, where
the whole projectile is absorbed, and the incomplete fusion, where only a part of the projectile charge is
absorbed by the target. As far as the total fusion is concerned, the cross section can be obtained
by using the scattering matrices. The CDCC total fusion cross sections
obtained in this way are shown in Fig.\ \ref{fig3} as red lines.
Above the Coulomb barrier, the present calculation overestimates the data by about a factor two.
This overestimation is due to the different definitions adopted for fusion in the experiment and in the calculation.
In other words, the theory defines fusion as the absorption of the whole projectile or 
any of its fragments, including the neutron. On the other hand, the experimental cross section
only includes contributions from the absorption of charge~\cite{DHB99}. Thus, the two cross sections
differ by the neutron absorption which, owing to the lack of a Coulomb barrier, may be quite large. 

The CDCC calculation can be adapted to the experimental conditions by using an equivalent 
definition  of the fusion cross section, and by removing the neutron-target imaginary potential.
The results obtained in this way are shown in Fig.\ \ref{fig3} as black lines. 
By excluding the neutron capture, we obtain an excellent agreement with experiment above $E_{\rm cm}\approx 40$ MeV. This is more clearly shown in the inset, drawn in a linear scale. 
We confirm previous CDCC calculations
\cite{DTB03,JPK14} which conclude that the influence of breakup channels increases at sub-barrier energies.
Below the Coulomb barrier, the single-channel calculation is in better agreement with experiment, and the full
calculation overestimates the data. This discrepancy
is common to most CDCC fusion calculations (see, for example, Fig. 1 of Ref.~\cite{JPK14}), and deserves further
investigation.

\section{Conclusion}
\label{sec5}
We have applied the CDCC formalism to the $\bepb$ system. In a first step, we have computed 
$\be$ wave functions in a three-body $\aan$ model. These calculations were performed for low-lying states, but also for pseudostates, which represent positive-energy approximations
of the continuum. The main advantage of the hyperspherical approach is that it treats the
three-body continuum without any approximation concerning possible $\ben$ or $\ahe$ cluster
structures, which are in fact strictly equivalent.

The three-body model of $\be$ relies on $\alpha+\alpha$ and $\alpha+n$ (real) interactions,
which reproduce very well the elastic phase shifts. With these bare interactions, the $\be$ ground state is slightly too bound. We therefore introduce a phenomenological three-body
force to reproduce the experimental ground-state energy. The spectroscopic properties of low-lying
states is in fair agreement with experiment. 

We used first the $\be$ wave functions in a CDCC calculation of the $\bepb$ elastic scattering
at energies close to the Coulomb barrier.  As expected, including continuum channels
improves the theoretical cross sections. A fair agreement with the data is obtained.

We also applied the four-body  CDCC to the $\bepb$ fusion, by using the same potentials
as for elastic scattering. The total fusion cross section is found in good agreement with
experiment, except at very low energies, where we overestimate the data.


\end{document}